\documentclass[12pt]{article}

\def\beq{\begin{equation}} \def\eeq{\end{equation}}
\def\bea{\begin{eqnarray}} \def\eea{\end{eqnarray}}

\def\beann{\begin{eqnarray*}} \def\eeann{\end{eqnarray*}}

\let\a=\alpha \let\be=\beta \let\g=\gamma \let\de=\delta
\let\e=\varepsilon \let\z=\zeta  
 \let\k=\kappa \let\la=\lambda 
\let\n=\nu    \let\s=\sigma
\let\om=\omega

\let\qd=\quad  

\let\Ts=\textstyle  

\def\0{\over } \def\1{\vec }     \def\2{{1\over2}} \def\4{{1\over4}}
\def\5{\bar }  \def\6{\partial } \def\7#1{{#1}\llap{/}}

\def\<{\langle } \def\>{\rangle }

\let\auf=\uparrow \let\ab=\downarrow

\def\i{{\rm i}} \def\tr{\mbox{tr}}

\begin{document}

\newfont{\elevenmib}{cmmib10 scaled\magstep1}%
\newcommand{\tabtopsp}[1]{\vbox{\vbox to#1{}\vbox to12pt{}}}
\font\larl=cmr10 at 32pt
\newcommand{\es}{\got s}
\newcommand{\st}{\phantom{\mbox{\Large I}}\hspace{-6pt}\relax}
\newcommand{\preprint}{
            \begin{flushleft}
   \elevenmib Yukawa\, Institute\, Kyoto\\
            \end{flushleft}\vspace{-1.3cm}
            \begin{flushright}\normalsize  \sf
            YITP-01-51\\
           {\tt hep-th/0106130} \\ June 2001
            \end{flushright}}
\newcommand{\Title}[1]{{\baselineskip=26pt \begin{center}
            \Large   \bf #1 \\ \ \\ \end{center}}}
\hspace*{2.13cm}%
\hspace*{0.7cm}%
\newcommand{\Author}{\begin{center}\large \bf
           V.\,I.\, Inozemtsev\footnote{
permanent address: BLTP JINR, 141980 Dubna, Moscow Region, Russia}
 and R.\, Sasaki \end{center}}
\newcommand{\Address}{\begin{center}
            Yukawa Institute for Theoretical Physics\\
     Kyoto University, Kyoto 606-8502, Japan
      \end{center}}
\newcommand{\Accepted}[1]{\begin{center}{\large \sf #1}\\
            \vspace{1mm}{\small \sf Accepted for Publication}
            \end{center}}
\baselineskip=20pt

\preprint
\thispagestyle{empty}
\bigskip
\bigskip
\bigskip

\Title{Scalar Symmetries of the Hubbard Models with 
Variable Range Hopping}
\Author

\Address
\vspace{1cm}

\begin{abstract}
Examples of scalar conserved currents are  presented for trigonometric,
hyperbolic and elliptic versions of the Hubbard model with non-nearest
neighbour variable range hopping. 
They support for the first time the hypothesis about the integrability
of the elliptic version. 
The  two-electron wave functions are  constructed
in an explicit form.

\end{abstract}
\bigskip
\bigskip
\bigskip



\newpage
 The Hubbard models
with non-nearest-neighbour hopping first proposed 
by Gebhard and Ruckenstein
[1-4] have received considerable attention 
since this  family of models itself contains the usual
Hubbard model on an infinite lattice as well 
as the Haldane-Shastry  [5,6]
 spin chain as
certain limiting cases, and so do the generators of its Yangian
symmetry [7]. Here we would like to present  some conserved
currents belonging to the scalar part 
of the algebra which provide an evidence for the integrability of more
 general
models with hopping given by elliptic functions.
In paper [7] it has been shown that trigonometric and hyperbolic
models with hopping matrices 
$t_{jk}=t(j-k)=\sin^{-1}{\pi\over N}(j-k)$
and $t_{jk}=t(j-k)=\sinh^{-1}\kappa (j-k)$ have an explicit 
Yangian symmetry which, however,
does not imply integrability. One needs also the scalar part of the
 whole symmetry algebra which has been found in [8] 
for the Haldane-Shastry
spin chain and in [9] for the conventional Hubbard model.

The basic model describes electrons of spin $\s=\pm1/2$ 
created by operators
$c_{j\s}^+$ at site $j$ of a one dimensional lattice. The
probability amplitude for hopping between sites $j$ and $k$ will be
denoted by $t_{jk}$.
The strength of the repulsive interaction of two electrons 
of different spin 
on the same lattice site is $U > 0$. In these notations, 
the Hamiltonian reads
\beq \label{ham}
     H = \sum_{j, k} t_{jk} c_{j \s}^+ c_{k \s} +
         2U \sum_j (c_{j \auf}^+ c_{j \auf} - {\Ts \frac{1}{2}})
                   (c_{j \ab}^+ c_{j \ab} - {\Ts \frac{1}{2}}).
\eeq
 We shall consider translational invariant hopping
amplitudes, $t_{jk} = t_{j - k}$. 

There is a canonical transformation, namely
\beq \label{cano}
     c_{j \ab} \rightarrow c_{j \ab}, \qd
     c_{j \auf} \rightarrow c_{j \auf}^+, \qd
     U \rightarrow - U.
\eeq
This transformation leaves every Hamiltonian of the form (1) with
antisymmetric hopping matrix invariant, but the global spin
operators and the Yangian generators are not invariant under (2).

To write down the $su_2$ generators of
the rotational symmetry of the Hamiltonian (\ref{ham}) and generators 
of other symmetries, it is  convenient to use
 spin operators formed as linear combinations of
products of one creation and one annihilation operator at one
 site. 
It is  also useful  to introduce spin-like
operators with indices corresponding to two different sites so as one
arranges the pair of operators $c_{j\s}^+ c_{k\tau}$ in a
$2 \times 2$-matrix labeled by spin indices $\s$ and $\tau$,
 $(S_{jk})_{\tau}^{\s} =
c_{j\s}^+ c_{k\tau}$, 
\beq
     S_{jk}^\a = \tr({\s}^{*\a} S_{jk}), \qd
     S_{jk}^0 = \tr(S_{jk}), \qd S_j^\a := S_{jj}^\a, \qd
     S_j^0 = S_{jj}^0,
\eeq
where the $\s^\a$ are the Pauli matrices.
 Note that $(S_{jk}^\a)^+
= S_{kj}^\a$, $(S_{jk}^0)^+ = S_{kj}^0$. 
The spin density and electron density operators are denoted by
$\frac{1}{2}S_j^\a$
and $S_j^0$, respectively.   
The commutators of these operators can be written explicitly,
\bea \label{c1}
     [S_{jk}^0,S_{lm}^0] & = &
         \de_{kl} S_{jm}^0 - \de_{mj} S_{lk}^0, \\[1ex]
     \label{c2}
     [S_{jk}^0,S_{lm}^\a] & = &
         \de_{kl} S_{jm}^\a - \de_{mj} S_{lk}^\a, \\[1ex]
     \label{c3}
     [S_{jk}^\a,S_{lm}^\be] & = &
         \de^{\a \be} \left(
         \de_{kl} S_{jm}^0 - \de_{mj} S_{lk}^0 \right)
         + \i \, \e^{\a \be \g} \left(
         \de_{kl} S_{jm}^\g + \de_{mj} S_{lk}^\g \right).
\eea
However, there are other bilinear relations due to the composite 
 nature of these operators.
The Hamiltonian (\ref{ham}) now takes the following form
\beq \label{eggs}
     H = \sum_{j,k} t_{jk} S_{jk}^0 + U \sum_j \left(
         (S_j^0 - 1)^2 - \Ts{\frac{1}{2}} \right).
\eeq
Since the particle number $I^0 = \sum_j S_j^0$ is
conserved, only the term $(S_j^0)^2$ is relevant in the
interaction part of the Hamiltonian 
while the other terms commute with $H$
and can be removed by a shift of the chemical potential.

 To provide examples of the conserved currents which might exist for 
some choice of the hopping matrix,
 consider the ansatz
\bea
J=\sum_{j\neq k}^{N}\left[A_{jk}S_{jk}^{0}+
B_{jk}(S_{j}^{0}S_{k}^{0}-\vec S_{j}
\vec S_{k})+D_{jk}(S_{j}^{0}+
S_{k}^{0})S_{jk}^{0}+E_{jk}(S_{jk}^{0})^2\right],
\eea
in which $N$ is the number of sites. 
The condition $[H,J]=0$ with the use of (4-6) can 
be cast into the form of two
functional equations
\bea
4t_{jk}(B_{lk}-B_{jl})+(t_{jl}D_{lk}-D_{jl}t_{lk})&=&0,
\\
2(t_{jk}E_{kl}+t_{kj}E_{jl})+(t_{jl}D_{kl}+t_{kl}D_{jl})&=&0.
\eea
The coefficients $A$ can be expressed in terms of $B$, $D$ and $E$ as
\bea
A_{jk}=-2D_{jk}+(2U)^{-1}
[-8t_{jk}B_{jk}+2t_{kj}E_{jk}-a_{jk}],
\eea
where 
\[ a_{jk}=\sum_{l\neq j ,k}
t_{jl}D_{lk}.
\]
Several boundary equations for $t$, $B$ 
and $D$ must be satisfied, too:
\begin{eqnarray*}
\sum_{l\neq j,k}(t_{jl}A_{lk}-A_{jl}t_{lk})&=&0,\\
\sum_{k\neq j}(t_{jk}D_{kj}-D_{jk}t_{kj})&=&0,\\
\sum_{k\neq j}(t_{jk}A_{kj}-t_{kj}A_{jk})&=&0.
\end{eqnarray*}
 The first functional equation
is just the Calogero-Moser one with known general 
analytic solution $\psi$:
\[
\psi(x)={{\s(x+\la)}\over{
\s(x)\s(\la)}}e^{\n x}.
\]
By using it,
one can express all the unknown structures in (9-11)
 (recall that $ t_{jk}=t(j-k) $ etc.) as follows, 
\[
t(x)=t_{0}\psi(x),\qquad B(x)=-{d\over4}\psi(x)\psi(-x),
\]
\[
D(x)=d\left[\psi'(x)-({{h\wp'(\la)}\over2}+
\z(\la)+\n)\psi(x)\right],\
E(x)={{d\psi^2(x)}\over2}\left[\st
1-h\psi(x\!+\!\la)\psi(-x\!-\!\la)\right],
\]
\[
a(x)=t_{0}d\psi(x)\!\left[\st -(N-3)\wp(x)+h_{1}(N-2)\xi(x)+
(\xi(x)-h_{1})
(2x\z(N/2)-N\z(x))+s\right],
\]
\[
\xi(x)=\z(x+\la)-\z(x)-\z(\la), \quad h_{1}=h\wp'(\la)/2,
\quad s=-(N-2)\wp(\la)-\sum_{l=1}^{N-1}\wp(l),
\]
where $\s$, $\z$ and $\wp$ are the Weierstrass elliptic 
functions determined by
the full periods $\om_{1}=N, \om_{2}=i\k$.
The other parameters are given by $\la=i\a$ or $i\a+N/2$, 
$\n=-2\zeta(N/2)
\la N^{-1}$, in which
 $\k$, $d$, $h$ and $\a$ are 
arbitrary real numbers. 
Besides this general solution, there are of course
the degenerate rational, trigonometric and hyperbolic 
forms which correspond
to one or two infinite periods of the Weierstrass function. 
The vanishing of
the boundary terms can be verified rather easily 
for the degenerate cases;
in general  elliptic case one needs to perform long but straightforward 
calculations which show that the boundary terms
also disappear at the above choices. 
The key formula in these calculations 
is the following identity
\begin{eqnarray*}
&& [\wp(y+\la)-\wp(\la)]\,[\z(x-y)-\z(x+\la)+\z(y)+\z(\la)]+\\
&& [\wp(x+\la)-\wp(\la)]\,[\z(y-x)-\z(y+\la)+\z(x)+\z(\la)]=\wp'(\la).
\end{eqnarray*}
 One thus can see that the formulas for $t$, $B$, $D$, $E$ and $
a$ define scalar conserved
 currents for two three-
parametric families of Hubbard models
defined by the sets
$(\la=i\a, \k, U/t_{0})$ and   $(\la=N/2 +i\a, \k, U/t_{0})$. 
 At the points $\la=\om_{1}/2,\om_{2}/2$
or $(\om_{1}+\om_{2})/2,$ 
i.e. at $\la$ being half-periods
of the Weierstrass function $\wp$, the function $\psi(x)$ becomes odd
and another  independent current appears under the canonical 
transformation
(2). It would be of interest to verify by direct calculations 
that both currents commute
as it takes place for the two copies of the 
Yangian operators in the case
of trigonometric and hyperbolic hopping [7]. 
One can thus conclude that
these three-parametric families of Hubbard models might be integrable.
 An  important
 open question is to confirm this conjecture 
by constructing the complete
 set   of conserved currents commuting with the 
Hamiltonian. One can also see that
the trigonometric and hyperbolic models of 
Bares, Gebhard and Ruckenstein
fall into these families if one of the periods of the 
Weierstrass function,
$\om_{1}$ or $\om_{2}$, tends to $\infty$ under 
an appropriate choice of
$\la$.

Let us now calculate exact two-electron wave functions for 
the Hamiltonian
(1) with the general elliptic hopping matrix. 
The problem is nontrivial
only for the $S=0$ sector of the model. 
The wave functions can be written as
\bea
\Phi=\sum_{j\neq k}\phi_{jk}c_{j\uparrow}^{+}c_{k\downarrow}^{+} 
\vert0\rangle+
\sum_{j}g_{j}c_{j\uparrow}^{+}c_{j\downarrow}^{+}\vert0\rangle,
\eea
where $\vert0\rangle$ is the vacuum state. The eigenequation
$H\Phi=E\Phi$ can be written in the form
\begin{eqnarray}
\sum_{s\neq j,k}(t_{js}\phi_{sk}+t_{ks}\phi_{js})\
&+&(t_{jk}g_{k}+t_{kj}g_{j})=E\phi_{jk},\\
\sum_{k\neq j}t_{jk}(\phi_{jk}+\phi_{kj})&=&(E-2U)g_{j}.
\end{eqnarray}
Note that $\phi_{jk}$ can always be decomposed into a sum
of the symmetric and antisymmetric parts
\[
\phi_{jk}={1\over 2}(\phi_{jk}+\phi_{kj})+{1\over 2}
(\phi_{jk}-\phi_{kj}).
\]
For antisymmetric $\phi_{jk}$, the equation (14) would be satisfied
trivially by $g_{j}=0$, and the solution to the equation (13)
would be given by antisymmetrized product of one-electron plane waves.
Thus from now on we concentrate on the nontrivial 
symmetric part, or in other words, assume that 
$\phi_{jk}$ is symmetric:
$$\phi_{jk}=\phi_{kj}.$$
 The ansatz for $\phi_{jk}$ and $g_{j}$ reads
\bea
\phi_{jk}=e^{i(p_{1}j+p_{2}k)}\varphi(j-k)
+e^{i(p_{1}k+p_{2}j)}\varphi(k-j),
\eea
$$
g_{j}= g_{0}e^{i(p_{1}+p_{2})j},
$$
where
\bea
\varphi(j)=\varphi_{0}{{\sigma(j+\tau)}\over {\sigma(j)}}.
\eea
The parameters $p_{1}$ and $p_{2}$ are connected with $\tau$ by the
conditions
of the periodicity of 
$\phi_{jk},$ $\phi_{j+N,k}=\phi_{j,N+k}=\phi_{jk},$
\bea
e^{ip_{1}N+\eta_{1}\tau}=1, \quad e^{ip_{2}N-\eta_{1}\tau}=1,
\eea
where $\eta_{1}=2\zeta(N/2)$.
The problem is to find all the parameters $p_{1,2}$,  
$g_{0}$ and $\tau$ from
the eigenequations (13-14) if the ansatz (15-16) is correct.
Recall that the elliptic hopping matrix in general is given by
\bea
t(j-k)=t_{0}\psi(j-k)=
t_{0}e^{\nu(j-k)}{{\sigma(j-k+\la)}\over{\sigma(j-k)\sigma(\la)}},
\eea
where the factor $\nu=-2\z (N/2)\la N^{-1}$ 
is chosen so as to satisfy the periodicity
condition, $t(j-k+N)=t(j-k)$. With the use of (16) and (18)
the second eigenequation (14) can be cast into the form
\bea
(E-2U)g_{0}=2t_{0}\varphi_{0}S_{1}(p_{1},p_{2},\tau),
\eea
where
$$S_{1}(p_{1},p_{2},\tau)=\sum_{s\neq 0}e^{\nu s}
{{\sigma(s+\lambda)}\over{\sigma^2 (s)\sigma(\la)}}
\left[\st e^{-ip_{2}s}\sigma(s+\tau)+e^{-ip_{1}s}
\sigma(s-\tau)\right].$$
The first eigenequation can be written as
\bea
F(\tau,p_{1},p_{2},j,k)+
F(\tau,p_{1},p_{2},k,j)+
F(-\tau,p_{2},p_{1},j,k)+
F(-\tau,p_{2},p_{1},k,j)
\eea
$$=-g_{0}\left[\st
t(j-k)e^{i(p_{1}+p_{2})k}+t(k-j)e^{i(p_{1}+p_{2})j}\right]
+E\phi_{jk},$$
where
$$F(\tau,p_{1},p_{2},j,k)=
\sum_{s\neq j,k}t(j-s)e^{i(p_{1}s+p_{2}k)}
\varphi(s-k)$$
$$=e^{i(p_{1}j+p_{2}k)}\sum_{q\neq 0,k-j}
t(-q)e^{ip_{1}q}\varphi(q+j-k).$$
Let us now calculate the sum in the last expression with the use of 
the explicit
forms of $t(j)$,  $\varphi(j)$ (16), (18), and introduce the notation
 \bea
{\cal S}(l)=\sum_{q\neq
0,l}^{N-1}{{\sigma(q-\lambda)\sigma(q-l+\tau)}\over{\sigma
(q)\sigma(\lambda)\sigma(q-l)}}e^{(-\nu+ip_{1})q},
\eea
where $l=k-j\in {\bf Z}$,
and the function of a continuous argument $x$,
\bea
G(l,x)=\sum_{q=0}^{N-1}{{\sigma(q-\lambda+x)
\sigma(q-l+\tau+x)}\over{\sigma
(q+x)\sigma(\lambda)\sigma(q-l+x)}}e^{(-\nu+ip_{1})q}.
\eea
The function $ G(l,x)$ is double quasiperiodic,
$$G(l,x+1)=e^{\nu-ip_{1}}G(l,x),\quad
G(l,x+\omega_{2})=e^{\eta_{2}(\tau-\lambda)}
G(l,x),$$
where $\eta_{1}=2\zeta(N/2)$ and  $\eta_{2}=2\zeta(\omega_{2}/2)$
 (recall that $\om_{2}$ is the second period of all 
the Weierstrass functions here). It has a simple pole
at $x=0$ with decomposition near it of the form
\begin{eqnarray*}
G(l,x)_{\vert x\to0}=
&-& {{\sigma(-l+\tau)}\over{\sigma(-l)}}\left(x^{-1}
+\zeta(-l+\tau)-\zeta(-l)
-\zeta(\lambda)\right) \\
&+& {{\sigma(l-\la)\sigma(\tau)}\over{\sigma(l)\sigma(\la)}}
\left(x^{-1}+\zeta(l-\la)-\zeta(l)+\zeta(\tau)
\right)e^{l(-\nu+ip_{1})}+{\cal
S}(l).
\end{eqnarray*}
 On the other hand, this function can be written in the form
$$G(l,x)=G_{0}(l)e^{rx}{{\sigma_{1}(x+\mu)}\over{\sigma_{1}(x)}},$$
where $\sigma_{1}$ is the Weierstrass sigma function with quasiperiods
$(1,\om_{2})$ and $G_{0}(l)$ is a constant factor. 
The parameters $r$ and $\mu$ can be found from the quasiperoidicity
conditions,
\begin{eqnarray*}
r(\tau,p_{1})&=&(2\pi i)^{-1}\left(\st
\eta_{12}(\nu-ip_{1})-
(\tau-\la)\eta_{2}\eta_{11}\right),\\
\mu(\tau, p_{1})&=&(2\pi
i)^{-1}\left(\st -\om_{2}(\nu-ip_{1})+(\tau-\la)\eta_{2}\right),
\end{eqnarray*}
where $\eta_{11}=2\zeta_{1}(1/2)$ and 
$\eta_{12}=2\zeta_{1}(\om_{2}/2)$.
Comparing the decompositions of both forms near 
$x=0$, one finds the sum
${\cal S}(l)$ explicitly,
\begin{eqnarray}
{\cal S}(l)&=&
-{{\sigma(-l+\tau)}\over{\sigma(-l)}}\left(\st
\zeta_{1}(\mu)+r-\zeta(-l+\tau)+
\zeta(-l)+\zeta(\lambda)\right) 
\\
&&+ {{\sigma(l-\la)\sigma(\tau)}\over{\sigma(l)\sigma(\la)}}
\left(\st
\zeta_{1}(\mu)+r-\zeta(l-\la)+\zeta(l)-\zeta(\tau)\right)
e^{l(-\nu+ip_{1})}.
\nonumber
\end{eqnarray}
The explicit form of $F(\tau,p_{1},p_{2},j,k)$ now reads with the use
of equations (22), (23) as follows,
\begin{eqnarray*}
&&F(\tau,p_{1},p_{2},j,k)\\
&&\qquad=
-{{\sigma(j-k+\tau)}\over{\sigma(j-k)}}
\varphi_{0}t_{0}e^{i(p_{1}j+p_{2}k)}
\!\left(\st\zeta_{1}(\mu)+r-\zeta(j-k+\tau)+
\zeta(j-k)+\zeta(\lambda)\right) \\
&&\qquad\quad +\ \varphi_{0}\sigma(\tau)t(j-k)e^{i(p_{1}+p_{2})k}
\!\left(\st\zeta_{1}(\mu)+
r+\zeta(j-k+\la)-\zeta(j-k)-\zeta(\tau)\right).
\end{eqnarray*}
Now, taking explicit summation of all four $F$'s with different
arguments in equation (20), one finds its compact form
\begin{eqnarray*}
&&- \phi_{jk}t_{0}
\left[\st\zeta_{1}(\mu)+\zeta_{1}(\tilde \mu)+r+\tilde r
 +2\zeta(\la)\right] \\
&&
+ \varphi_{0}\sigma(\tau)\left[\st 
t(j-k)e^{i(p_{1}+p_{2})k}+t(k-j)e^{i(p_{1}+p_{2})j}\right]
\left[\st \zeta_{1}(\mu)-\zeta_{1}(\tilde \mu)+r-\tilde r 
-2\zeta(\tau)\right]\\
&&=E\phi_{jk}-g_{0}
\left[\st t(j-k)e^{i(p_{1}
+p_{2})k}+t(k-j)e^{i(p_{1}+p_{2})j}\right], 
\end{eqnarray*}
where
$\tilde r=r(-\tau,p_{2})$, $\tilde \mu=\mu(-\tau, p_{2})$.
Comparing the coefficients in both sides, one obtains two equations
\begin{eqnarray}
 E&=&-t_{0}
\left[\st \zeta_{1}(\mu)+\zeta_{1}(\tilde \mu)+r+\tilde r
 +2\zeta(\la)\right], 
\\
g_{0}&=&-\varphi_{0}\sigma (\tau)
\left[\st \zeta_{1}(\mu)-\zeta_{1}(\tilde \mu)+r-\tilde r
 -2\zeta(\tau)\right].
\end{eqnarray}
These equations define $E$ and $g_{0}/\varphi_{0}$ in terms of $p_{1},
p_{2}$ and $\tau$. Plugging them into (19) results in the equation
for determining the phase shift $\tau$. 
Together with the relations between $\tau,p_{1}$ and $p_{2}$ (17)
 this equation allows one to determine all the parameters of the
 ansatz (14) for two-electron wave functions.

To summarize, we have found new scalar conserved 
currents for the Hubbard 
model with an elliptic hopping matrix and its 
trigonometric and hyperbolic degenerations. We proposed
the ansatz for the two-electron $S=0$ wave function and proved that it
allows
one to determine  momenta and phase shift of two-electron states. 
It would
be of interest to investigate the question of the completeness of our
 solution. In the case of the Heisenberg chain, it is known that the
analogous 
ansatz gives complete description of all two-magnon states. 
At this time,
it is not clear how to generalize this ansatz for the case of three or
more electrons, and to find higher conserved currents, but our results
give clear indications for the integrability of elliptic families of
the 1D Hubbard models.

{\it Acknowledgments}. V.I.I. is supported by JSPS long term fellowship.
R.S. is partially supported by the Grant-in aid from the Ministry of 
Education, Culture, Sports, Science and Technology, Japan, priority
area (\#707) "Supersymmetry and unified theory of elementary particles".
 
{\bf References}  
\begin{enumerate}
\item
F.~Gebhard and A.~E. Ruckenstein, Phys. Rev. Lett. {\bf 68}, 244 (1992).

\item
P.-A. Bares and F.~Gebhard, Europhys. Lett. {\bf 29}, 573 (1995);
			    J. Phys.  {\bf C7}, 2285 (1995).

\item
F.~Gebhard, A.~Girndt, and A.~E. Ruckenstein, Phys. Rev. 
{\bf B49}, 10926 (1994).

\item
D.~F. Wang, Q.~F. Zhong, and P.~Coleman, Phys. Rev.  {\bf B48},
8476 (1993).
\item

F.~D.~M. Haldane, Z.~N.~C. Ha, J.~C. Talstra, D.~Bernard, and
V.~Pasquier, Phys. Rev. Lett. {\bf 69}, 2021 (1992).

\item
D.~Bernard, M.~Gaudin, F.~D.~M. Haldane, and V.~Pasquier, J. Phys. 
{\bf A26}, 5219 (1993).

\item
F. G\"ohmann and V. Inozemtsev. Phys. Lett.  {\bf A214} 161 (1996).

\item
J.~C. Talstra and F.~D.~M. Haldane.
J. Phys. {\bf A28}, 2369 (1995).

\item
B.~S. Shastry, J. Stat. Phys. {\bf 50}, 57 (1988).

\end{enumerate}

\end{document}